# Enhanced Predictive Modeling for Hazardous Near-Earth Object Detection: A Comparative Analysis of Advanced Resampling Strategies and Machine Learning Algorithms in Planetary Risk Assessment

Sunkalp Chandra[1]

[1]Columbia University
New York, NY 10027
sc5895@columbia.edu

**Abstract**

This study evaluates the performance of several machine learning models for predicting hazardous near-Earth objects (NEOs) through a binary classification framework, including data scaling, power transformation, and cross-validation. Six classifiers were compared, namely Random Forest Classifier (RFC), Gradient Boosting Classifier (GBC), Support Vector Classifier (SVC), Linear Discriminant Analysis (LDA), Logistic Regression (LR), and K-Nearest Neighbors (KNN). RFC and GBC performed the best, both with an impressive F2-score of 0.987 and 0.986, respectively, with very small variability. SVC followed, with a lower but reasonable score of 0.896. LDA and LR had a moderate performance with scores of around 0.749 and 0.748, respectively, while KNN had a poor performance with a score of 0.691 due to difficulty in handling complex data patterns. RFC and GBC also presented great confusion matrices with a negligible number of false positives and false negatives, which resulted in outstanding accuracy rates of 99.7% and 99.6%, respectively. These findings highlight the power of ensemble methods for high precision and recall and further point out the importance of tailored model selection with regard to dataset characteristics and chosen evaluation metrics. Future research could focus on the optimization of hyperparameters with advanced features engineering to further the accuracy and robustness of the model on NEO hazard predictions.

**Introduction**

Asteroids, known more precisely as near-Earth objects, or NEOs, are a set of celestial bodies that constantly orbit around the sun coming dangerously close in path to the earth's surface. These bodies however present both scientific possibilities and threats to human life if not properly monitored. Due to new developments in the fields of space exploration and monitoring of how space objects behave in modern times, it is of great importance to learn the impact of an object in space. Research of NEOs has resulted in the accumulation of significant knowledge on their natures which enables Earth authorities to estimate their orbits and physical natures. A vital secondary objective of this research work relates to the investigation of the characteristics that make the asteroids classified as potentially hazardous asteroids.

This research in turn focuses on analysing the set of objects that NASA collected in its NeoWs program. The NeoWs API consists of a variety of information including the physical characteristics of a particular NEO, its orbital parameters, and even information pertaining to its close approaches. These studies focused on understanding the major characteristics and factors contributing to the wearing of potentially harmful asteroids. The research question focuses on the use of machine learning algorithms to identify potentially hazardous and non-hazardous asteroids.

Traditional methodologies for the identification of hazardous asteroids have relied mostly on observational techniques and manual classifications based on a very limited set of parameters, such as size, velocity, and proximity to Earth. While these approaches were effective up to a certain point, they lacked scalability and precision for the amount of asteroids monitored today. With the increasing availability of data-driven methodologies, machine learning has emerged as a potential tool to uncover patterns and predict classifications within large datasets. Past research has identified the capability of machine learning algorithms in handling large volumes of data in astronomy, especially in the identification of anomalies and classification of celestial objects.

This research is important for planetary defense and space exploration. Recently, there has been increased awareness about the threat of asteroids. Increased international efforts in detection, tracking, and mitigation strategies for asteroid impacts are underway. Detection of an asteroid that poses a hazard will represent an essential initial step in undertaking an Earth-protection strategy from any possible collisions. Not only planetary defense, but understanding the physical and orbital properties of asteroids contributes to the development of space mining, resource uses, and further solar system exploration.

Asteroids have also very important contributions to forming the outline of our understanding of the universe. Impacts by asteroids may involve massive events in Earth's history, including mass extinctions and the delivery of organic materials for initiating life. These space bodies act paradoxically while also contributing importantly to critical infrastructure in systems such as satellites and communication all over the world, this situation calls for the necessity of detailed monitoring and prediction.

The hypothesis that grounds this research is that machine learning algorithms classify asteroids as hazardous or nonhazardous based on a mix of their physical and orbital features. The scientific phenomenon to be investigated here is the dynamic behavior of asteroids when interacting with Earth's orbit, whose deviations in either velocity or trajectory may signal closeness in proximity. Coupled with state-of-the-art analytical techniques, this will help connect these phenomena for a better understanding of factors that make an asteroid potentially hazardous.

The goal of this paper is the classification of asteroids as hazardous or not, using machine learning models based on their features. Predictions involve identifying such patterns in the data that, for instance, the diameter and velocity of an asteroid and its close approach distance will give a good insight into the probability of its being hazardous. Large-diameter asteroids with low close approach distances might have higher labels as hazardous. The model's performance will be evaluated based on metrics such as accuracy, precision, recall, and F1 score to ensure the most reliable classification.

This paper, therefore, presents an experimental approach towards feature selection, model training, and validation using the NeoWs dataset. Predictors include physical features like diameter or absolute magnitude, and orbital ones such as semi-major axis or eccentricity. The response variable can be hazardous or not, depending on the asteroid under consideration.

**Materials**

The primary material needed for this research was a computer. For this specific research study, the computer that was used was a 2023 M2 MacBook Air with a 15.4-inch screen size. This computer was chosen for its power and because of the efficiency of the macOS operating system.

The programming language used for this study was Python (version 3.12.1). Python is widely used for various applications, particularly machine learning. Popular libraries in Python such as NumPy, Pandas, and Scikit-Learn were all imperative for this machine-learning research.

The study also required the use of Google Colaboratory which is a cloud-based platform that supports Python programming for machine learning tasks. Colaboraty provided a virtual machine that has GPU capabilities and this eliminated the need for local hardware as the processing capabilities of Google's servers can be used instead.

The dataset used in this research was found on the website: https://cneos.jpl.nasa.gov/.

**Data Preprocessing**

The raw dataset sourced from the NeoWs API required serious preprocessing to be reliable for any usable analysis. Physical features range from estimated diameters, while orbital parameters range from semi-major axis to eccentricity. After the preprocessing steps had been considered, the very crucial step involved missing value handling. Numerical features are missing value-imputed by their means or medians so that the data distribution remains representative, and the mode was imputed for the categorical variable.

Normalization and standardization are two of the techniques used in the preprocessing steps to prepare machine learning models with the data. Normalization scaled all features like relative velocity and miss distances in the range [0,1] using the following formula:

$$x' = \frac{x - x_{\min}}{x_{\max} - x_{\min}}$$

This can ensure that no feature or factor is dominated by other features with larger ranges while contributing to a model. In features related to orbital eccentricity, standardization would transform values into a standard normal distribution using the following formula in order to reduce the impact of outliers and improved model convergence:

$$z = \frac{x - \mu}{\sigma}$$

Feature engineering was employed to derive new attributes, enriching the dataset. For example, one such feature extracted could be a hazard probability feature derived on the basis of proximity and size thresholds. The categorical ranges for the estimated diameters allowed stratification for analyses that aided a better understanding of asteroid classifications.

**Exploratory Data Analysis**

Extensive exploratory data analysis (EDA) provided a deep dive into the structure and relations within the dataset. Visualizations like histograms and density plots of asteroid size, velocity, and miss distance showed the distribution of their values. These graphs outlined a pattern in the way diameters or velocities cluster around a particular value.

Relationships between the key variables-for example, how miss distances affect the hazardous classification of asteroids-were considered by using scatter plots. To quantify such relationships, Pearson correlation coefficients were computed:

$$\rho(X,Y) = \frac{\text{cov}(X,Y)}{\sigma_X \sigma_Y}$$

This analysis emphasized dependencies among variables, such as the strongly negative correlation that existed between estimated diameter and miss distance, influencing the hazardous classification.

**Machine Learning Models**

Different machine learning models were implemented for the classification of asteroids into hazardous and non-hazardous classes and the prediction of key parameters. The machine learning models used for this project have been described in detail below.

**Logistic Regression**

Logistic regression was a baseline model for binary classification. It predicts the probability of an asteroid being hazardous using the logistic function:

$$P(Y = 1|X) = \frac{1}{1 + e^{-(\beta_0 + \beta_1 X_1 + \ldots + \beta_n X_n)}}$$

Features such as estimated diameter and miss distance were used as inputs, providing a simple yet effective starting point.

**Random Forest**

The Random Forest algorithm is an ensemble model, which captures non-linear relationships by aggregating predictions from multiple decision trees:

$$\hat{f}(X) = \frac{1}{M} \sum_{m=1}^{M} T(X; \Theta_m)$$

This was quite effective for feature importance analysis, to show which variables were most influential.

**Support Vector Classifier**

Support Vector Classifiers were used for the classification of asteroids by identifying a hyperplane that separates hazardous and non-hazardous classes optimally:

$$\operatorname{argmin}_{w,b} \left( \frac{1}{2} ||w||^2 + C \sum_{i=1}^{n} \max(0, 1 - y_i(w \cdot x_i + b)) \right)$$

Their kernel functions enabled them to manage complex boundaries of decisions.

**Gradient Boosting Models**

Gradient Boosting, specifically XGBoost, performed the best for imbalanced classification to optimize a custom loss function:

$$\text{Obj} = \sum_{i=1}^{n} l(\hat{y}_i, y_i) + \sum_{j=1}^{T} \Omega(T_j)$$

Additionally, in the formula above $\Omega(T_j)$ regularized tree complexity. The incorporation of regularization terms reduced overfitting and enhanced generalization.

**Neural Networks**

Neural networks played an important role in this study. Feedforward neural networks learn complex nonlinear patterns through layers of weighted connections:

$$y = f(W_2 f(W_1 X + b_1) + b_2)$$

Recurrent Neural Networks and their variant, Gated Recurrent Units, modeled sequential dependencies in the data. The GRU architecture with update gates is defined as:

$$z_t = \sigma(W_z x_t + U_z h_{t-1} + b_z)$$

This was particularly effective in the analysis of temporal patterns, such as time changes in asteroid trajectories.

**Gaussian Naive Bayes**

Gaussian Naive Bayes (GaussianNB) classified the asteroids as hazardous or non-hazardous based on probabilistic modeling of the features. This model works on Bayes' theorem with the assumption of feature independence in each class. GaussianNB models the likelihood of each feature using a Gaussian distribution, which makes it suitable for continuous data such as asteroid diameter, relative velocity, and orbital eccentricity.

The posterior probability of a class $C$ given the input $x$ is calculated as:

$$P(C|x) = \frac{P(x|C)P(C)}{P(x)}$$

For GaussianNB, the likelihood *P(x|C)* for a feature *x* is modeled as:

$$P(x|C) = \frac{1}{\sqrt{2\pi\sigma_C^2}} \exp\left(-\frac{(x-\mu_C)^2}{2\sigma_C^2}\right)$$

Where $\mu_C$ and $\sigma_C^2$ are the mean and variance of the feature in class $C$

In this project, GaussianNB provided a simple but effective way of classifying asteroids, taking advantage of its probabilistic framework for handling features like estimated diameter and relative velocity. It is efficient and thus very good for initial comparisons, although it might face problems when the independence of features is violated.

**Linear Discriminant Analysis**

Linear Discriminant Analysis, or LDA, is used to classify asteroids into hazardous and non-hazardous objects by finding the linear combinations of features that best separate the two classes. The main assumption in LDA is that each class is drawn from a Gaussian distribution with a shared covariance matrix but differing mean vectors. Under this assumption, LDA can obtain decision boundaries that maximize class separability. The decision rule in LDA can be derived from the following formula for the discriminant function:

$$\delta_k(x) = x^T \Sigma^{-1} \mu_k - \frac{1}{2}\mu_k^T \Sigma^{-1} \mu_k + \log P(C_k)$$

Where:

$\delta_k(x)$ is the discriminant score for class $k$,

$\mu_k$ is the mean vector for class $k$,

$\Sigma$ is the shared covariance matrix across all classes,

$P(C_k)$ is the prior probability of class $k$.

LDA maximizes the ratio of between-class variance to within-class variance for optimal separation of classes. This was very useful for understanding linear separability in features like miss distances and orbital parameters. Though its performance was limited by the assumption of normally distributed data, interpretable results were obtained with much insight provided into the structure of the data set.

**Gaussian Process Classifier (GPC)**

GPC was used because it can model complex, nonlinear decision boundaries in the asteroid classification task. The GPC is a Bayesian model that defines a distribution over functions and uses it to classify data points probabilistically. It is highly flexible, allowing the classification of hazardous asteroids even in the presence of intricate relationships among features. The classification decision is based upon the posterior distribution over the possible functions $f(x)$, given the observed data:

$$P(y|x, \mathcal{D}) = \int P(y|f(x))P(f(x)|\mathcal{D})df(x)$$

Where:

$P(y|x, \mathcal{D})$ is the posterior probability of class $y$,

$P(y|f(x))$ is the likelihood of class $y$ given the function value $f(x)$,

$P(f(x)|\mathcal{D})$ is the posterior distribution over functions given the dataset $\mathcal{D}$.

The kernel function, $K(x, x')$, plays a crucial role in defining the similarity between data points. Popular choices include the Radial Basis Function (RBF) kernel:

$$K(x, x') = \exp\left(-\frac{\|x - x'\|^2}{2\ell^2}\right)$$

where $\ell$ is the length scale parameter.

In this project, GPC provided a non-linear approach to classify asteroids based on complex interactions among features like orbital inclination, perihelion distance, and relative velocity. Its probabilistic outputs were particularly useful for uncertainty quantification, though its computational complexity limited scalability for larger datasets.

**KNeighborsClassifier (KNN)**

The KNeighborsClassifier classified the asteroid based on the k-nearest neighbors in the feature space. This nonparametric algorithm relied on a majority vote of the nearest neighbors for the determination of a point class.

The decision rule for a point $x$ is:

$$y = \mathrm{argmax}_c \sum_{i=1}^{k} I(y_i = c)$$

Where:

$y$ is the predicted class for $x$,

$I$ is the indicator function, returning 1 if $y_i = c$ and 0 otherwise,

$k$ is the number of nearest neighbors considered.

The distance metric, typically Euclidean, is defined as:

$$d(x, x') = \sqrt{\sum_{i=1}^{n}(x_i - x'_i)^2}$$

In the context of this project, KNN provided a simple and effective way to classify hazardous asteroids based on proximity to similar data points in feature space. While it did well on smaller datasets, it struggled with large-scale data due to its computational inefficiency.

**Model Evaluation Metrics**

The performance of each model was assessed using a combination of the following quantitative metrics. These metrics ensured robust evaluation, particularly for the imbalanced nature of hazardous and non-hazardous classes. Confusion matrices also provided a visual representation of the prediction accuracy.

**Accuracy:**

$$\text{Accuracy} = \frac{\text{TP} + \text{TN}}{\text{TP} + \text{TN} + \text{FP} + \text{FN}}$$

**Precision:**

$$\text{Precision} = \frac{\text{TP}}{\text{TP} + \text{FP}}$$

**Recall:**

$$\text{Recall} = \frac{\text{TP}}{\text{TP} + \text{FN}}$$

**F-1 Score:**

$$F1 = 2 \cdot \frac{\text{Precision} \cdot \text{Recall}}{\text{Precision} + \text{Recall}}$$

**F2 Score:**

$$F_2 = \frac{(1 + 2^2) \cdot \text{Precision} \cdot \text{Recall}}{(2^2 \cdot \text{Precision}) + \text{Recall}}$$

**Model Evaluation with Cross-Validation**

Various algorithms were implemented and their performance in classifying asteroid hazards was evaluated using Repeated Stratified K-Fold Cross-Validation. This helps ensure the stability of the model's performance and makes it independent of any particular split, while preserving the distribution of both classes in each fold. The models evaluated included Linear Discriminant Analysis (LDA), Random Forest Classifier (RFC), Gradient Boosting Classifier (GBC), K-Nearest Neighbors (KNN), Support Vector Classifier (SVC), and Logistic Regression (LR). Each model was evaluated using the F2 score, a metric that emphasizes recall, making it ideal for the asteroid classification task where detecting hazardous asteroids is critical. The ranking methodology consisted of building a Pipeline for each model, embedding into it the preprocessing steps: MinMaxScaler for feature scaling and PowerTransformer to make data more Gaussian-like. This way, all features would be more or less within the same range, and models can learn complicated patterns. The cross-validation was done using RepeatedStratifiedKFold, where the dataset was split into 10 and repeated 3 times to get a robust estimate of the performance of each model. Models were compared according to the mean and standard deviation of the F2 score, providing insights about the best algorithms that perform well in predicting asteroid hazard status. Additionally, confusion matrices and classification reports were also created in order to evaluate the performance of the models.

**Undersampling Methods for Class Imbalance**

In this work, some undersampling techniques were applied to the class imbalance in the asteroid hazard classification task. The main goal was balancing the dataset by reducing instances from the majority class, using different undersampling methods implemented in the imblearn library. The methods range from TomekLinks, which removes borderline majority class instances, to the EditedNearestNeighbours (ENN) which eliminates instances of the majority class misclassified by their nearest neighbors, its iterative version. RepeatedEditedNearestNeighbours (RENN), the OneSidedSelection (OSS) algorithm, which uses TomekLinks and ENN in combination for more aggressive cleaning, to the NeighbourhoodCleaningRule (NCR), in which both classes are cleaned by removing noisy instances from both the majority and minority classes. These under-sampling techniques were then incorporated into an imblearn pipeline, where MinMaxScaler and PowerTransformer were fitted for feature scaling and transformation to ensure that the data was properly normalized and more Gaussian-like, which improves model performance. Models were evaluated using RandomForestClassifier combined with cross-validation, specifically RepeatedStratifiedKFold (10 folds, 3 repeats), to ensure stable performance while preserving class distribution. Each model's performance was evaluated by the F2 score, which prioritizes recall since the identification of hazardous asteroids is very critical. The mean and standard deviation of the F2 score for each undersampling method were computed and compared for valuable insights into how each method influenced model performance.

**Oversampling Techniques for Class Imbalance**

In this research, oversampling techniques were employed to overcome the problem of class imbalance in asteroid hazard classification. It is aimed at balancing the dataset by generating synthetic instances of the minority class, which here are the potentially hazardous asteroids. Three oversampling techniques were utilized from the imblearn library: SMOTE- Synthetic Minority Over-sampling Technique, BorderlineSMOTE, and ADASYN- Adaptive Synthetic Sampling Approach. SMOTE works in the following way: It generates synthetic samples along the line segments that connect the available minority class samples, thus increasing the representation of the minority class. BorderlineSMOTE is an extension to SMOTE, specifically creating synthetic samples near the decision boundary, which serves to help models improve accuracy on more difficult instances to classify. ADASYN further refines this by focusing more on generating synthetic samples for the minority class instances that are harder to classify, i.e., those misclassified by the nearest neighbor classifier. These oversampling methods were integrated into an imblearn pipeline with MinMaxScaler for feature scaling and PowerTransformer for making feature distributions more Gaussian-like, improving learning processes for the models. First, the training of the models was performed with a RandomForestClassifier. Performance evaluation was done by Repeated Stratified K-Fold with 10 splits and 3 repeats for obtaining a robust estimate of model performance while keeping the class distribution. For this, the F2 score puts more emphasis on recall, which is something very important in hazardous asteroid detection. The mean and standard deviation of the F2 scores of every oversampling method were calculated and compared to find out which technique significantly enhances the model in predicting hazardous asteroids.

**Combined Sampling Techniques for Class Imbalance**

This project also adopted combined sampling methods to handle the class imbalance in asteroid hazard classification by integrating both oversampling and undersampling techniques. Two combined sampling methods, SMOTETomek and SMOTEENN from the imblearn library, were adopted. The SMOTETomek technique combines SMOTE (Synthetic Minority Over-sampling Technique) with an undersampling method called TomekLinks that removes instances of the majority class that are near the decision boundary. This combination aims to generate synthetic samples for the minority class while cleaning the majority class of noise, improving model performance. In SMOTEENN, SMOTE is combined with Edited Nearest Neighbors, where SMOTE produces synthetic instances for the minority class and ENN cleans the generated instances along with the actual data by removing misclassified instances based on their neighbors. These combined techniques were then incorporated into an imblearn pipeline with MinMaxScaler to feature scale and PowerTransformer for making features more Gaussian-like, which helps in improving the model's learning capability. For model training, the robust RandomForestClassifier was used as it is a well-known ensemble method that aggregates the predictions over several decision trees. The evaluation was done using Repeated Stratified K-Fold Cross-Validation, with 10 splits and 3 repeats, ensuring that the class distribution was maintained and performance was consistently measured. The F2 score was chosen as the evaluation metric, with a greater emphasis on recall, critical for the detection of hazardous asteroids. Hence, the performance of these combined sampling methods in terms of mean and standard deviation was calculated for the F2 score and compared.

# Results

Figure 1: Correlation Heatmap of Asteroid Orbital and Physical Parameters from Dataset

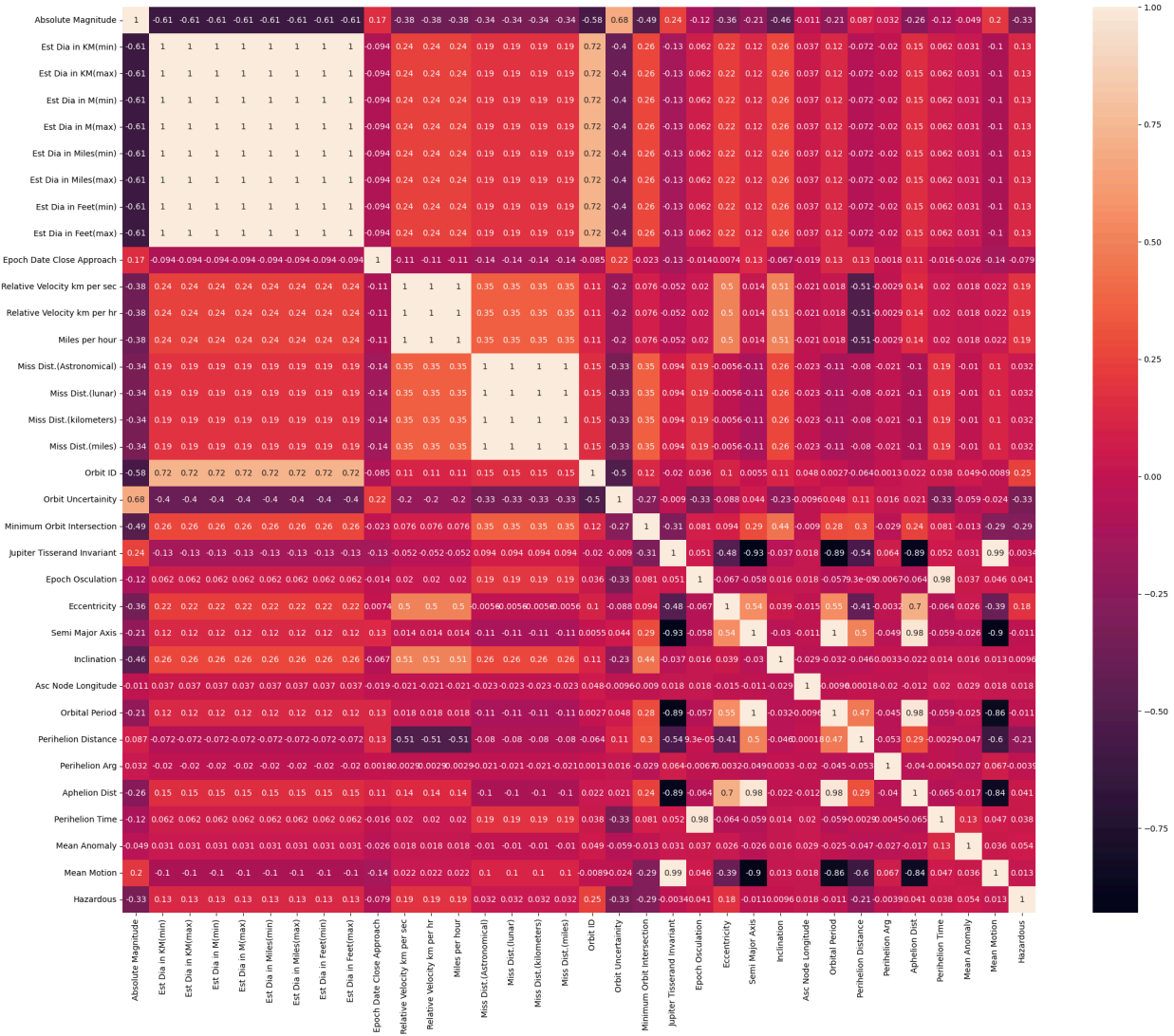

This heatmap provides a complete representation of the correlation matrix of the dataset under study. Each cell in the heatmap represents the Pearson correlation coefficient between two variables, ranging from -1 to 1. Lighter shades of orange represent positive correlations, while darker shades, moving toward purple, represent negative correlations. Values near to 1 reflect a strong positive relationship, while values near to -1 indicate a strong negative relationship. Diagonal elements have a perfect correlation (1.0) because each variable is correlated with itself.

Interestingly enough, some clusters of variables show a high inter-correlation level. For example, the estimated diameters in kilometers, meters, and feet are perfectly correlated (r ≈ 1), as they are simply proportional to each other by unit conversion. Likewise, strong correlations exist between the variables that describe miss distance in various units: astronomical units, lunar distances, and kilometers. In contrast, there are weak or negligible correlations between parameters like eccentricity and hazardous status, reflecting little direct linear relationship among these variables. It means this analysis gives important insight into variable relationships, further aiding feature selection and its interpretation in subsequent modeling.

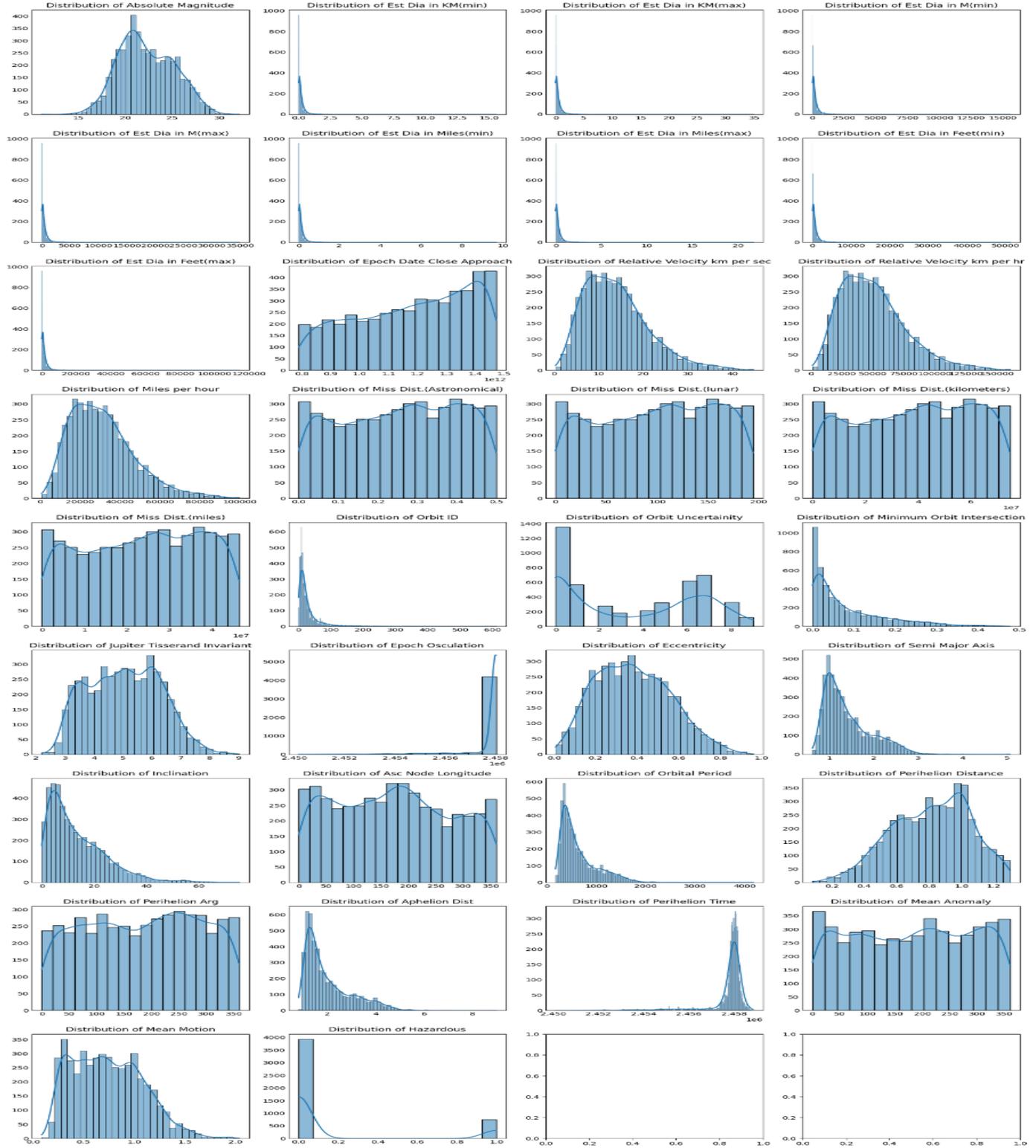

Figure 2: Distribution Plots of Asteroid Orbital and Physical Parameters

These plots above give a full insight into visualizing a large variety of physical and orbital parameters of the asteroid data set. The estimated diameter values in different units are hugely right-skewed. Most of the estimated diameter values are small in magnitude, and the frequency sharply decreases as the diameter value increases. All of these distributions have a tail due to a few outliers having significantly larger diameters, hinting at the existence of some quite big, rare asteroids. These types of skews reinforce the point of log scaling in normalizing these features for statistical modeling and machine learning.

Parameters connected to the dynamics of close approaches include relative velocity in km/s and km/hr, and miss distance in astronomical units, lunar distances, kilometers, and miles, each exhibiting a range of behaviors. The distribution of relative velocity is log-normal-like, with most asteroids traveling at moderate speeds and a few much faster. The miss distances are similarly spread, reflecting the wide range of asteroid trajectories relative to Earth. This uniformity in those variables suggests significant variability in the proximity of asteroids while approaching, which is an important variable in determining their potential risk.

Orbital properties give more insight into deeper interpretations and classification of asteroids. Orbit uncertainty, a measure of how precisely an asteroid's orbit is determined, is skewed to the right. This indicates that while most asteroids have well-determined orbits, a subset exhibits higher uncertainties, potentially due to limited observational data or complex orbital perturbations. Eccentricity, which quantifies the deviation of an orbit from being circular, peaks around 0.5, reflecting that many asteroids have moderately elliptical orbits. While the semi-major axis shows the size of the orbit, the inclination represents the tilt of the orbit relative to the plane of the solar system. The pattern for inclination is bimodal, suggesting distinct groups of asteroids: those whose orbits are near the ecliptic plane and those whose orbits are much more inclined, possibly as a result of dynamic interactions with other bodies.

Orbital period and distance parameters, like orbital period and perihelion distance, also show unique characteristics. The orbital period shows peaks that are consistent with common asteroid families due to gravitational influence from Jupiter. Similarly, the perihelion distance is all over the map, reflecting the various orbital paths that asteroids take as they approach their closest distance to the Sun. The perihelion argument distribution, ranging over all angles, stands to reflect the varying orientation of asteroid orbits within the solar system. The hazardous classification is a binary variable that is highly imbalanced, with most asteroids classified as non-hazardous. This is an important consideration for predictive modeling since any machine learning model will need to take into account the under-representation of hazardous asteroids. Features such as epoch osculation and ascending node longitude are periodic or uniformly distributed, which aligns with expectations based on orbital mechanics and the celestial coordinate systems used to describe asteroid orbits.

Figure 3: Boxplot Analysis of Asteroid Orbital and Physical Parameters

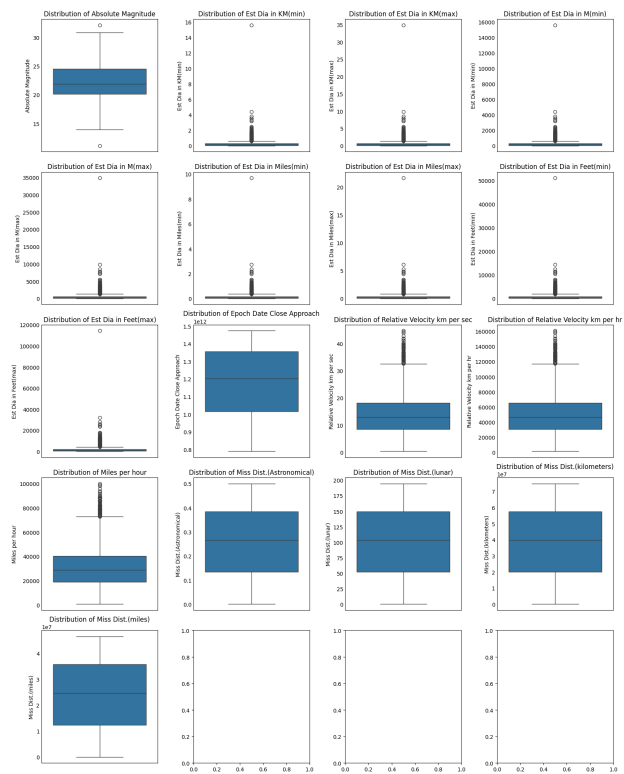

The figure presents boxplots that summarize the statistical distribution of asteroid orbital and physical parameters, including measures of central tendency, spread, and potential outliers.

Each boxplot visually represents the interquartile range (IQR) of a feature, with the box boundaries corresponding to the 25th and 75th percentiles, the central line indicating the median, and whiskers extending to 1.5 times the IQR. Points beyond the whiskers denote outliers.

The estimated diameters across various units, such as kilometers, meters, miles, and feet, are highly concentrated within a narrow range toward the lower end, with several outliers extending far beyond the whiskers. These correspond to a few very large asteroids, which accentuate the positive-skewed nature of these features. Although the scaling of the units is different, the shape of the boxplots is the same, reinforcing their proportional relationships.

The close approach parameters show large dispersions, such as relative velocity and miss distances. The range of values of the relative velocity in km/s, km/hr, and miles per hour is very large, with a great number of high-speed outliers. These high velocities indicate that the trajectories are hyperbolic or fast-moving near-Earth objects. The miss distance by astronomical units, lunar distances, and kilometers also spreads uniformly and does not show extreme skewness: whiskers cover a really wide range of distances. These suggest the great diversity of asteroid trajectories, which may be important for defining the levels of potential impacts on Earth. The epoch date of the close approach, in numeric date format, is tightly clustered with little deviation, indicating that there is a consistent time frame when recorded observations were taken. Eccentricity and orbit inclination should have variations to be indicative of the different orbital shapes and orbital tilts of the asteroid population. However, inclinations with outliers indicate a good number of asteroids are observed with unusual orbits possibly because of perturbations or other dynamic interactions.

Table 1: Distribution of Hazardous Asteroids

| Hazardous Classification | Count | Percentage |
|:---:|:---:|:---:|
| Non-hazardous (0) | 3,932 | 83.9% |
| Hazardous (1) | 755 | 16.1% |
| Total | 4,687 | 100% |

Table 1 summarizes the hazardous classification of asteroids in the dataset. The "Hazardous" variable is a binary feature, taking on either a value of 0 for non-hazardous or 1 for hazardous. Of the total 4,687 asteroids, 3,932 (83.89%) are classified as non-hazardous, and 755 (16.11%) are hazardous.

This notable imbalance in the dataset highlights that the majority of observed asteroids are not considered a significant risk to Earth. However, the 16.11% classified as hazardous is a critical subset, as these asteroids may pose potential impact threats and warrant closer monitoring.

This class imbalance, from a statistical and machine learning perspective, is huge. Predictive models trained on such data may develop a bias toward the majority class, that is, nonhazardous, unless some remedial measures such as oversampling of the minority class or undersampling of the majority class are taken, or some advanced techniques like class-weighted models are employed. The representation also stresses that hazardous asteroids should have greater priority in observational and analytical studies to enhance early detection and mitigation efforts.

Table 2: Summary Statistics of Asteroid Physical Metrics

| Metric | Absolute Magnitude Est | Dia in KM (min) Est | Dia in KM (max) Est | Dia in M (min) Est | Dia in M (max) Est | Dia in Miles (min) Est |
| --- | --- | --- | --- | --- | --- | --- |
| Count | 4687.00000 | 4687.00000 | 4687.00000 | 4687.00000 | 4687.00000 | 4687.00000 |
| Mean | 22.267865 | 0.204604 | 0.457509 | 204.604203 | 457.508906 | 0.127135 |
| std | 2.890972 | 0.369573 | 0.826391 | 369.573402 | 826.391249 | 0.229642 |
| min | 11.160000 | 0.001011 | 0.002260 | 1.010543 | 2.259644 | 0.000628 |
| 25% | 20.100000 | 0.033462 | 0.074824 | 33.462237 | 74.823838 | 0.020792 |
| 50% | 21.900000 | 0.110804 | 0.247765 | 110.803882 | 247.765013 | 0.068850 |
| 75% | 24.500000 | 0.253837 | 0.567597 | 253.837029 | 567.596853 | 0.157727 |
| max | 32.100000 | 15.579552 | 34.836938 | 15579.55241 | 34836.93825 | 9.680682 |

The table below gives an overall view of the summary statistics in terms of the key physical metrics used for asteroids: absolute magnitude and estimated diameter in kilometers, meters, and miles. The absolute magnitude ranges between 11.16, the more major and reflective the brighter objects, and 32.10 for the smaller and less bright ones, with an average value of 22.27. This is a clear indication that this database comprises mostly small-sized objects. The brightness variability puts into perspective the diversity of the asteroids, from small and faint to larger and brighter bodies.

Estimates of diameters put into view the physical scale of the asteroids in kilometers, meters, and miles, by minimum and maximum values. Diameters in kilometers range from as small as 0.001 km to as large as 34.84 km, while average minimum and maximum values are 0.204 km and 0.458 km, respectively. In meters, diameters range between 1.01 and 15,579.55 meters, with average minimum and maximum diameters being 204.60 and 457.51 meters, respectively. Similarly, diameters in miles range from 0.0006 to 9.68 miles with mean values of 0.13 miles minimum and 0.28 miles maximum. These statistics show great variability as obtained by the large standard deviations and the range of values obtained. This dataset is positively skewed due to a few large-sized asteroids since the maximum values obtained are far greater compared to the 75th percentile.

Table 3: Continued Summary Statistics of Asteroid Physical Metrics

| Metric | Dia in Miles (max) Est | Dia in Feet (min) Est | Dia in Feet (max) | Epoch Date Close Approach | Inclination | Asc Node Longitude |
|---|---|---|---|---|---|---|
| Count | 4687.00000 | 4687.00000 | 4687.00000 | 4687.00000 | 4687.00000 | 4687.00000 |
| Mean | 0.284283 | 671.273653 | 1501.013521 | 1.179881e+12 | 13.373844 | 172.157275 |
| std | 0.513496 | 1212.511199 | 2711.257465 | 1.981540e+11 | 10.936227 | 103.276777 |
| min | 0.001404 | 3.315431 | 7.413530 | 7.889472e+11 | 0.014513 | 0.001941 |
| 25% | 0.046493 | 109.784247 | 245.485039 | 1.015574e+12 | 4.962341 | 83.081208 |
| 50% | 0.153954 | 363.529809 | 812.877364 | 1.203062e+12 | 10.311836 | 172.625393 |
| 75% | 0.352688 | 832.798679 | 1862.194459 | 1.355558e+12 | 19.511681 | 255.026909 |
| max | 21.646663 | 51114.018738 | 114294.42050 | 1.473318e+12 | 75.406667 | 359.905890 |

This table summarizes the key metrics related to asteroid diameters in miles and feet, epoch dates of close approaches, and orbital parameters like inclination and ascending node longitude. In the case of diameter in miles, the maximum value observed is 21.65 miles, and the mean diameter is 0.28 miles. The big standard deviation of 0.51 miles underlines the fact that there is much variation in asteroid sizes, reflecting the diversity in asteroid populations. The diameter in feet varies from as low as 3.31 feet to as high as 114,294.42 feet, with averages of 671.27 feet minimum and 1501.01 feet maximum. The great range, combined with high variability, points to a few very large asteroids that skew the dataset.

The epoch date of close approach in Julian time has a mean of about $1.18 \times 10^{12}$ with a standard deviation of $1.98 \times 10^{11}$. This shows the temporal dispersion of the close approaches within the dataset. The minimum value is $7.89 \times 10^{11}$, while the maximum is $1.47 \times 10^{12}$, showing that the dataset covers a wide period for asteroid observation and prediction.

The inclination of asteroids' orbits, which is the tilt of their orbital planes concerning the ecliptic, ranges from 0.01° to a steep 75.41°, with an average inclination of 13.37°. This shows that most of the asteroids have low orbital inclinations typical of NEOs, while a few outliers have highly inclined orbits. The ascending node longitude, which defines the orientation of the orbit in space, spans nearly the entire 360° range (from 0.002° to 359.91°). The mean value of 172.16° indicates a uniform distribution of orbital nodes across the dataset.

Table 4: Summary Statistics of Orbital Dynamics

| Metric | Orbital Period | Perihelion Distance | Perihelion Arg | Aphelion Dist | Perihelion Time | Mean Anomaly | Mean Motion |
|---|---|---|---|---|---|---|---|
| Count | 4687.0000 | 4687.000 | 4687.0000 | 4687.000 | 4687.0000 | 4687.0000 | 4687.0000 |
| Mean | 635.582076 | 0.813383 | 183.93215 | 1.987144 | 2.457728e+06 | 181.167927 | 0.738242 |
| std | 370.954727 | 0.242059 | 103.51303 | 0.951519 | 9.442264e+02 | 107.501623 | 0.342627 |
| min | 176.557161 | 0.080744 | 0.006918 | 0.803765 | 2.450100e+06 | 0.003191 | 0.086285 |
| 25% | 365.605031 | 0.630834 | 95.625916 | 1.266059 | 2.457815e+06 | 87.006918 | 0.453289 |
| 50% | 504.947292 | 0.833153 | 189.76164 | 1.618195 | 2.457973e+06 | 185.718889 | 0.712946 |
| 75% | 794.195972 | 0.997227 | 271.77757 | 2.451171 | 2.458108e+06 | 276.531946 | 0.984669 |
| max | 4172.23134 | 1.299832 | 359.99309 | 8.983852 | 2.458839e+06 | 359.917991 | 2.039000 |

This table presents detailed statistical metrics on the orbital dynamics of asteroids, considering the orbital period, perihelion and aphelion distances, perihelion argument, perihelion time, mean anomaly, and mean motion. These parameters are of great relevance to understanding the trajectories and orbital behavior of asteroids.

Its orbital period is 635.58 days on average, or how much time an asteroid uses for one revolution around the Sun; with a standard deviation of about 370.95 days, some had shorter orbital periods of approximately 176.56 days, reaching a length as long as 4,172.23 days. Such a big standard deviation shows a serious dispersion of this characteristic among these asteroids, which might even belong to the inner solar system and rather long-orbit asteroids.

The perihelion distance is the point of the orbit closest to the Sun and has an average of 0.81 AU with a range from 0.08 to 1.30 AU. A low standard deviation of 0.24 AU supports that most asteroids in this data set follow orbits relatively close to Earth's vicinity. On the other hand, the average aphelion distance is 1.99 AU point farthest from the Sun-ranging from 0.80 to 8.98 AU. This is a broader range but with a higher standard deviation of 0.95 AU and reflects the presence of highly eccentric orbits among the asteroids.

The perihelion argument, which defines the orientation of perihelion in the orbital plane, spans nearly the full angular range from 0.007° to 359.99°, with a mean value of 183.93° and a large standard deviation of 103.51°, which indicates a wide distribution of perihelion positions among the asteroids due to their highly diversified orbital orientation.

The perihelion time, which is the time when the asteroid is closest to the Sun, centers around the mean value of $2.457 \times 10^6$ Julian days with a rather low variability-standard deviation of 944.23. This narrow range shows that this dataset captures a specific timeframe of asteroid perihelion events.

The mean anomaly is the position of the asteroid in its orbit at any given time, averaging 181.17° with values ranging from 0° to 359.91° almost uniformly. This even distribution reflects no significant clustering of asteroids in specific orbital positions. The mean motion, representing the rate of orbital progress, averages 0.74° per day, ranging from 0.086° to 2.04° per day. The higher rates correspond to asteroids with shorter orbital periods, as expected.

Together, these metrics offer critical insights into a variety of orbital properties of the asteroids-from tightly bound orbits with small perihelion distances to extended, eccentric paths that stretch much further into the solar system. This will be highly useful for making accurate predictions of asteroid trajectories, judging potential Earth-impact risks, and comprehending NEO dynamics.

Table 5: Table of Entropy Values of Asteroid Physical Metrics

| Metric | Absolute Magnitude Est | Dia in KM (min) Est | Dia in KM (max) Est | Dia in M (min) Est | Dia in M (max) Est | Dia in Miles (min) Est |
| --- | --- | --- | --- | --- | --- | --- |
| Entropy | 7.008145 | 7.008145 | 7.008145 | 7.008145 | 7.008145 | 7.008145 |

This table shows the calculated entropies of some key physical metrics of the asteroids, which include absolute magnitude, estimated diameter in kilometers minimum and maximum, and the estimated diameter in meters and miles. Entropy is a measure of uncertainty or disorder; it describes quantitatively the amount of unpredictability in the distribution of the metric under consideration. The higher entropy value reflects greater variability, and vice versa.

All the values of entropy, for all the listed metrics of absolute magnitude and estimated diameters in kilometers, meters, and miles (minimum and maximum), are 7.008145. This consistency across all metrics suggests that the distribution of these physical characteristics is similarly uncertain or disordered. The entropy value here is 7.008145, showing a rather high variability in the size and magnitude of the asteroids in this dataset, hence no clustering or coherence in these measurements.

These high entropy values show that the physical properties of the asteroids in this dataset, represented by absolute magnitude and size measurements, are highly varied within a broad range. This can be understood to reflect the diverse nature of the asteroid population, where the sizes and magnitudes of asteroids are highly different, with no clear pattern or trend across the dataset.

Table 6: Continued Table of Entropy Values of Asteroid Physical Metrics

| Metric | Dia in Miles (max) Est | Dia in Feet (min) Est | Dia in Feet (max) | Epoch Date Close Approach | Inclination | Asc Node Longitude |
|---|---|---|---|---|---|---|
| Entropy | 7.008145 | 7.008145 | 7.008145 | 9.434939 | 11.723491 | 11.723491 |

This table shows the continued entropy values for additional asteroid physical metrics, including maximum diameter in miles, minimum and maximum diameters in feet, and orbital parameters such as epoch date of close approach, inclination, and longitude of the ascending node. As mentioned earlier, entropy is a measure of disorder or uncertainty in the distribution of these metrics. Higher entropy values reflect greater variability or uncertainty in the dataset.

The entropy values for the maximum diameter in miles, and minimum and maximum diameters in feet are 7.008145, consistent with the values observed in the above table. This consistency suggests that these metrics also are highly variable like the physical size measures, and there is no strong central tendency or clustering.

In the orbital parameters of epoch date of close approach, inclination, and longitude of ascending node, the values of entropy are significantly higher. The entropy value for the epoch date of close approach is 9.434939, while that for inclination and ascending node longitude stands at 11.723491. This higher entropy value in the two other orbital characteristics of the dataset asteroids signals greater variability in the same. Higher entropy in these parameters suggests that the asteroids in this study are more diverse in their orbital characteristics, with large variability in their orbital inclinations and the timing of their closest approach to Earth.

These higher orbital parameter entropy values indicate increased complexity and less predictability in asteroid orbits. With changing inclinations and differences in epochs of close approach, the simple and predictable pattern for asteroids could not be manifested in their orbits, further manifesting heterogeneity within the asteroid population.

Table 7: Table of Entropy Values of Orbital Dynamics

| Metric | Orbital Period | Perihelion Distance | Perihelion Arg | Aphelion Dist | Perihelion Time | Mean Anomaly | Mean Motion |
|---|---|---|---|---|---|---|---|
| Entropy | 11.723491 | 11.723491 | 11.723491 | 11.72349 | 11.723491 | 11.723491 | 11.723491 |

Table 7 gives the entropy values of orbital dynamics: orbital period, perihelion distance, perihelion argument, aphelion distance, perihelion time, mean anomaly, and mean motion. The entropy value of these orbital metrics is the same as 11.723491, showing that they all are highly variable and unpredictable for the asteroid dataset. These relatively equal entropies further suggest that the orbital nature of the asteroids is highly mixed and uncertain, showing very complicated and diverse orbital behavior amongst objects.

Figure 4: Accuracies of All Models for the Prediction of Hazardous NEOs

| Model | Accuracy |
|---|---|
| DecisionTreeClassifier | 0.995736 |
| RandomForestClassifier | 0.995736 |
| GradientBoostingClassifier | 0.995736 |
| LogisticRegression | 0.948827 |
| SVC | 0.943497 |
| GaussianNB | 0.928571 |
| LDA | 0.902985 |
| GPC | 0.900853 |
| KNeighborsClassifier | 0.892324 |

The performance evaluation of different classification algorithms proved to be quite informative about their accuracy. DecisionTreeClassifier, RandomForestClassifier, and GradientBoostingClassifier were the best models, with an accuracy of 99.57%. This demonstrates that these are really strong models for classifying this dataset and may be suitable for similar data. The Logistic Regression model had a respectable accuracy of 94.88%, showing that it is reliable to achieve reasonably high performance in this context.

The rest of the models showed specific performances, where SVC had an accuracy rate of 94.34% and GaussianNB with an accuracy of 92.85%. LDA and GPC showed somewhat lower performance with 90.29% and 90.08%, respectively. The lowest performance was recorded by KNeighborsClassifier with 89.23%.

Figure 5: Model Performance Comparison Based on F2-Score with Cross-Validation

```
>LDA 0.749 (0.047)
>RFC 0.987 (0.015)
>GBC 0.986 (0.014)
>KNN 0.691 (0.052)
>SVC 0.896 (0.026)
>LR 0.748 (0.054)
```

The cross-validation performance of various machine learning models, evaluated with the F2-score, has shown significant differences. Among these, the best performance was given by the RFC, which had a mean F2-score of 0.987 with a low standard deviation of 0.015, thus combining high accuracy with stability of the model performance across different folds. Similarly, the GBC showed an F2-score of 0.986 with a standard deviation of 0.014, which proves its reliability for the dataset. These results show how powerful ensemble methods are in capturing the complex patterns of the data.

Other models showed different performances. The SVC achieved a reasonably high F2-score of 0.896 with a standard deviation of 0.026, followed by LDA and LR, which yielded respective mean F2-scores of 0.749 and 0.748. However, with their higher standard deviations at 0.047 and 0.054, respectively, they are showing some variability in their performance. The KNeighborsClassifier had the smallest F2-score of 0.691 with the biggest standard deviation of 0.052, which makes it the least consistent among the models. These observations pinpoint the importance of model selection and tuning based on the dataset at hand and the metric of evaluation.

Figure 6: Confusion Matrix for RandomForestClassifier - Accuracy and F2-Score Analysis

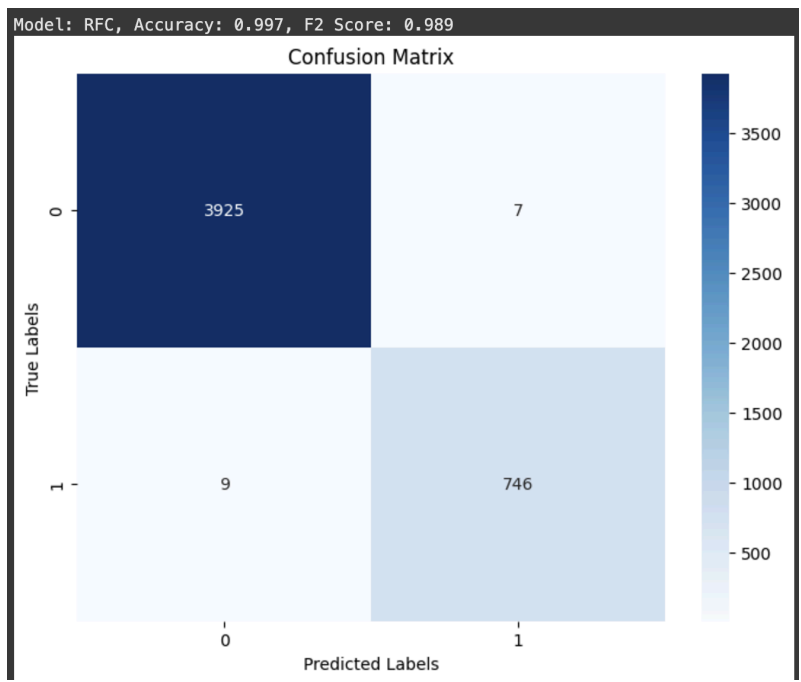

The confusion matrix depicts the performance of the classification by the RFC model, which had an accuracy of 99.7% and an F2 score of 0.989. From the confusion matrix, it can be obtained that the model correctly classified 3,925 instances of class 0 and 746 instances of class 1, hence having high precision and recall for both classes. This is further supported by the minimal misclassifications (7 false positives and 9 false negatives). These results highlight the strength of the RandomForestClassifier in differentiating between the two classes with minimum errors. From the high F2 score of the model, it is well suited for situations where recall must be weighted more than precision in a model.

Figure 7: Confusion Matrix for GradientBoostingClassifier - Accuracy and F2-Score Performance

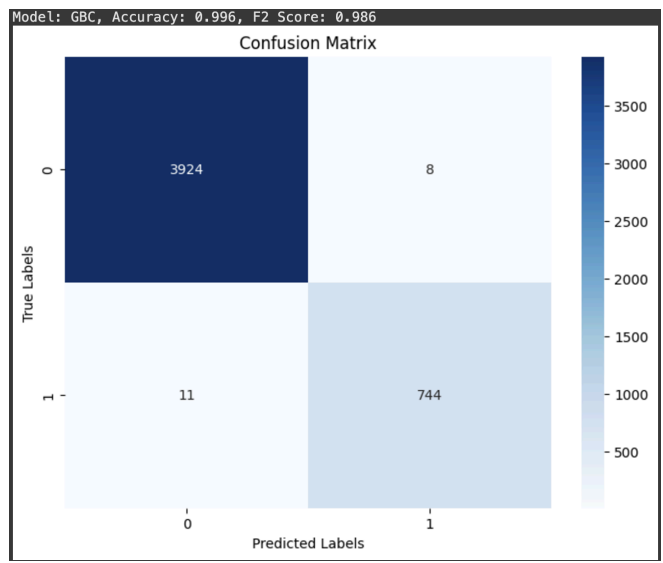

The confusion matrix of the GBC model shows an excellent performance that achieved an accuracy of 99.6% and an F2 score of 0.986. It rightly classified 3,924 instances of class 0 and 744 instances of class 1. This result highlights the strong capability of the model to balance precision and recall effectively, especially in those cases where minimizing false negatives is crucial.

Figure 8: Confusion Matrix for SupportVectorClassifier - Accuracy and F2-Score Performance

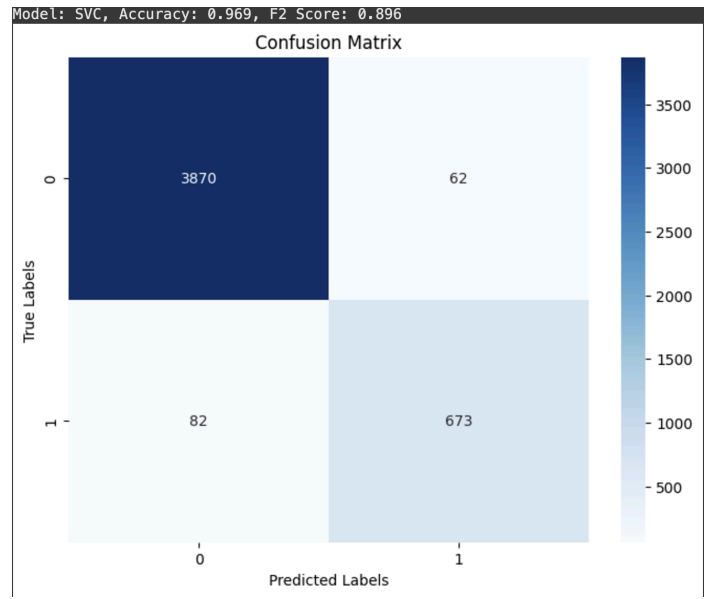

The confusion matrix of the SVC model shows a slightly worse performance that achieved an accuracy of 96.9% and an F2 score of 0.896. However, this result still highlights the ability of the model to balance precision and recall effectively.

Figure 9: Performance Comparison of Oversampling Techniques Using F2-Score with Cross-Validation

```
>SMOTE 0.988 (0.013)
>BLS 0.992 (0.010)
>ADASYN 0.991 (0.011)
```

Different oversampling techniques were performed, including SMOTE, BorderlineSMOTE (BLS), and ADASYN, and their performance was measured with the pipeline that included data scaling, power transformation, and a RandomForestClassifier. Among the different oversampling methods, BorderlineSMOTE (BLS) showed the best performance and reached an average F2-score of 0.992 with a standard deviation of 0.010, hence showing both accuracy and stability. ADASYN followed closely with an average F2-score of 0.991 and a standard deviation of 0.011, showing its effectiveness in handling class imbalance while holding on to consistent performance. Meanwhile, SMOTE reached a 0.988 F2-score with a slightly higher standard deviation of 0.013, reflecting a marginally weaker but still strong performance. These results showed that oversampling techniques enhance classifier performance on imbalanced datasets and that BLS has a slight advantage with respect to mean F2-score and consistency.

Figure 10: Performance Comparison of Undersampling Techniques Using F2-Score with Cross-Validation

```
>TL 0.987 (0.012)
>ENN 0.991 (0.011)
>RENN 0.989 (0.010)
>OSS 0.989 (0.014)
>NCR 0.989 (0.009)
```

The performance of different under-sampling techniques has been explored in this work, including Tomek Links (TL), Edited Nearest Neighbours (ENN), Repeated Edited Nearest Neighbours (RENN), One-Sided Selection (OSS), and Neighbourhood Cleaning Rule (NCR), using a pipeline that included data scaling, power transformation, and a RandomForestClassifier. ENN reached the best F2-score of 0.991 with a standard deviation of 0.011, reflecting its superior handling of class imbalance while maintaining stability. RENN, OSS, and NCR were close behind with the same average F2-score of 0.989, with NCR having the lowest standard deviation of 0.009, indicating its consistency across cross-validation folds. TL was slightly lower, with an F2-score of 0.987 and a standard deviation of 0.012, which still showed very good performance.

Figure 11: Performance Comparison of Combined Sampling Techniques Using F2-Score with Cross-Validation

```
SMOTE—TOMEK 0.990 (0.013)
SMOTE + KKN 0.991 (0.008)
```

The performance of combined sampling methods SMOTE-TOMEK and SMOTE + KKN was evaluated by using a RandomForestClassifier encapsulated within the preprocessing pipeline. The highest F2-score of 0.991 for SMOTE + KKN was associated with a very small standard deviation of 0.008, which, besides its good score, pointed toward stability across folds. Then comes SMOTE-TOMEK with an F2-score of 0.990 and a standard deviation of 0.013, hence showing a little more variation but still robust.

**Discussion**

This study investigates various oversampling, undersampling, and hybrid techniques to address class imbalance in the prediction of hazardous near-Earth objects (NEOs), a critical task in planetary defense and space science. Given the highly imbalanced nature of NEO datasets, where the majority of objects are non-hazardous and a small fraction are considered potentially hazardous, the goal was to optimize the performance of machine learning classifiers in identifying these hazardous objects. The models were evaluated by several resampling techniques that would enhance the classifier for the detection of those rare but critical events and reduce the number of false negatives. The base model used was the RandomForestClassifier, and F2-scores and accuracy were used to assess the performance to pinpoint the trade-off between precision and recall, with the main focus on improving the recall for hazardous NEOs.

Methods of oversampling first employed to generate synthetic data belonging to the minority class were based on SMOTE, Borderline-SMOTE, and ADASYN. BLS worked out the best, yielding the highest F2-score of 0.992 with the lowest variability of 0.010; thus, being the most stable and efficient technique in raising the recall of the classifier. This probably stems from the fact that BLS has focused on generating synthetic samples from the hard borderline instances, which are more likely to be misclassified, and because it enhances the model to be better at identifying hazardous NEOs. This is further supported by the strong performance of ADASYN (F2 = 0.991, SD = 0.011) regarding targeted oversampling of harder-to-classify instances for enhancing the detection of hazardous NEOs. Although still powerful, SMOTE only achieved an F2-score of 0.988 with higher variability, SD = 0.013; it generates valid synthetic data but performance may vary depending on the nature of the decision boundary.

This is particularly important in the case of imbalanced data, such as that of hazardous NEO prediction, where the cost of failing to predict a hazardous object is very high. These methods balance the dataset by synthesizing data points that mimic the minority class, thus increasing the classifier's sensitivity to the minority class and its ability to predict potentially hazardous NEOs more accurately.

The methods of undersampling lessen the number of instances belonging to the majority class (that is, non-hazardous NEOs). Amongst the undersampling techniques, Edited Nearest Neighbours outperformed the rest with an F2-score of 0.991 and a standard deviation of 0.011. ENN works by removing those instances of the majority class that are misclassified by their nearest neighbors, yielding a cleaner decision boundary that improves the recall for the minority class. This is important in the identification of hazardous NEOs because noisy or overlapping instances of the majority class can mask the true boundary between hazardous and non-hazardous objects. Repeated Edited Nearest Neighbours (RENN), One-Sided Selection (OSS), and Neighbourhood Cleaning Rule (NCR) all showed similar results (F2 = 0.989), with only slight differences in performance stability. These methods remove instances from the majority class based on various heuristics designed to clean the decision boundary. Tomek Links (TL) performed worse, F2 = 0.987, probably because it removes only pairs of instances that are near each other but from different classes, which may not be enough when the decision boundaries in NEO datasets are more complex. These under-sampling methods also aim at balancing the data distribution by focusing on the removal of those instances of the majority class that do not contribute to the model's ability to distinguish the minority class (hazardous NEOs). The low variance obtained for methods such as ENN and NCR suggests that these approaches lead to more stable models, especially in the context of identifying hazardous NEOs, where stability is crucial to avoid false negatives.

SMOTE-TOMEK and SMOTE + KKN combined methods were tried as hybrid techniques to handle oversampling and undersampling simultaneously. SMOTE-TOMEK (F2 = 0.990, SD = 0.013) couples the SMOTE with the Tomek Links approach for generating synthetic samples in the minority class while performing the cleaning of the decision boundary by removing noisy instances from the majority class. This approach reduces overfitting with minimal loss of important information on hazardous NEOs, and it is a good candidate for further applications in the field. In the same vein, the best performance among combined techniques was given by SMOTE + KKN with F2 = 0.991, SD = 0.008. In other words, the SMOTE strategy of generating synthetic data and KNN in the process of cleaning allowed a fine-tuning decision boundary in SMOTE+KKN with fewer false positives and reduced numbers of false negatives in hazard neo-identifications. Beyond this, low standard deviation provides great strength to this algorithm in practice and makes it a very hopeful choice for future neo-detectors. These hybrid techniques are of particular value in tasks like NEO hazardous detection, where both the accurate generation of minority class samples and refinement of the decision boundary are crucial for reducing the risk of overlooking dangerous objects. This ability to combine the strengths of both oversampling and undersampling methods leads to more stable models that can accurately identify hazardous NEOs with minimal risk of overfitting.

     Although accuracy is often used as a general measure of model performance, in the context of predicting hazardous NEOs, F2-scores provide a more appropriate evaluation metric due to the imbalanced nature of the dataset. A high accuracy could easily be achieved by always predicting the majority class, non-hazardous NEOs, but this would not help detect hazardous NEOs. For example, even though SMOTE had a good score in accuracy, this oversampling technique performed pretty average on the F2-score because it predicted the majority class but still failed to return dangerous NEOs. In the inverse case, techniques that can have higher F2-scores such as BLS, ADASYN, and SMOTE + KKN demonstrated improved recall for hazardous NEOs because this is a very serious predictive factor in the forecast of a dangerous object in the environment of space. The low variance observed in methods such as SMOTE + KKN and BLS further indicates that these techniques are not only effective in improving F2 scores but also provide stable and consistent results across different validation sets, which is essential in ensuring the reliability of NEO detection systems.

     This study provides a comprehensive comparison of various oversampling, undersampling, and hybrid techniques for the task of predicting hazardous near-Earth objects, yet several avenues remain for future improvement and exploration. While the RandomForestClassifier was used in this work, the use of

other ensemble methods, like Gradient Boosting Machines or XGBoost, might further improve the models. These methods are usually more robust against overfitting and may result in even better performance in identifying hazardous NEOs. Future work could consider deep learning models, such as CNNs, which have proven to be very successful in image classification tasks, or RNNs, which can be adapted for time-series or sequence-based NEO data. These models may find more complex patterns and interactions in the data than traditional classifiers can. Feature enhancement for predicting hazardous NEOs could increase the performance. Further studies can be taken up with orbital characteristics, velocity, or proximity to Earth, apart from other data. Integrating more detailed data may let the model better estimate the real risk that a NEO can produce. Due to the challenges in obtaining large labeled datasets of hazardous NEOs, one future work is to explore the use of transfer learning to leverage pre-trained models on similar datasets (e.g., astronomical data from other sources) to improve predictive performance. Finally, future work could take the line of real-time prediction of hazardous NEOs using techniques such as SMOTE-TOMEK and SMOTE + KKN incorporated into a continuous monitoring system that will flag potential hazards as new data arrives.

In conclusion, the results of this study show that advanced resampling techniques such as BLS, SMOTE + KKN, and SMOTE-TOMEK contribute to a remarkable increase in the performance of machine learning models on hazardous NEO detection. The class imbalance problem is mitigated, enabling higher recall and, therefore, enabling better identification of dangerous NEOs which is a key requirement for planetary defense activities.